\documentclass[12pt]{IOPART}
\usepackage{graphicx}

\newcommand{\ltsim}{\mbox{{\raisebox{-0.4ex}{$\stackrel{<}{{\scriptstyle\sim
}}
$}}}}

\def\hc2{$H_{c2}$}

\def\nh4{$\alpha$-(BEDT-TTF)$_2$NH$_4$Hg(SCN)$_4$}
\def\sruo{Sr$_{2}$RuO$_{4}$}
\begin{document}

\title[Cyclotron resonance in \sruo]
{Cyclotron resonance in the layered perovskite superconductor
\sruo}
\author{Eva Rzepniewski$^1$,
Rachel S. Edwards$^1$, John Singleton$^1$, Arzhang Ardavan$^1$
and Y.~Maeno$^2$.}

\address{$^1$University of Oxford, Department of Physics, The
Clarendon Laboratory, Parks Road,
Oxford OX1 3PU, U.K.}
\address{$^2$Department of Physics, Kyoto University, Kyoto 606-8502, Japan}

\begin{abstract}
We report a detailed study of the
magnetic-field-orientation dependence
of the millimetre-wave magnetoconductivity of
the superconductor \sruo .~
We find two harmonic series of cyclotron resonances.
We assign the first, corresponding to
a quasiparticle mass of $4.29 \pm 0.05~m_{\rm e}$,
where $m_{\rm e}$ is the free-electron mass,
to the $\beta$ Fermi-surface section.
We assign the second series, which contains only odd harmonics,
to cyclotron resonance of the $\gamma$ Fermi-surface section,
yielding a quasiparticle mass of $12.35 \pm 0.20 ~m_{\rm e}$.
A third, single cyclotron resonance,
corresponding to a quasiparticle mass of
$5.60 \pm 0.03 m_{\rm e}$, is attributed
to the $\alpha$ Fermi-surface section.
In addition, we find a very strong absorption mode
in the presence of a magnetic field component parallel
to the quasi-two-dimensional planes of the sample.
Its dependence on the orientation of the magnetic field
cannot be described in the context of conventional cyclotron resonance,
and the origin of this mode is not yet clear.
\end{abstract}

\submitto{\JPC}
\maketitle
\section{Introduction}
Great interest surrounded the
discovery of superconductivity in \sruo ~in 1994~\cite{discovery3}.
The structural similarity of \sruo ~to
``high-$T_{\rm c}$'' cuprate superconductors along with the striking
differences in superconducting transition temperature
(La$_{2-x}$Ba$_{x}$CuO$_{4}$ has $T_{\rm c}\approx 30$~K whereas
Sr$_{2}$RuO$_{4}$ has $T_{\rm c}\approx 1.4$~K) invited comparisons
and further investigation~\cite{maeno-summary3}.
Since then, compelling evidence for the existence of an
unconventional spin-triplet superconducting state in
\sruo ~has been found~\cite{maeno-summary3}.
Large quasiparticle mass enhancements~\cite{mackenzie-ARPES3} suggest
that strong electron correlation effects, favoring unconventional
superconductivity, are important in
\sruo ~\cite{discovery3,maeno-summary3}. Another key
signature of unconventional superconductivity--- the strong
suppression of $T_{\rm c}$ by non-magnetic as well as magnetic
impurities--- is also found~\cite{defect3},
and Josephson junction experiments point to non s-wave
pairing~\cite{Josephson3}. Nuclear magnetic resonance (NMR) and
nuclear quadrupole resonance (NQR) experiments show  behaviour
which cannot be explained by the s-wave model~\cite{NMR3}: the
relaxation rate $1/T_{\rm 1}$ drops abruptly below $T_{\rm c}$
without the Hebel-Slichter coherence peak, and the relation
$T_{\rm 1}T=$constant remains true well below $T_{\rm c}$.

Rice and Sigrist suggested
the possibility of an odd-parity ($l=1$, p-wave) pairing state in
Sr$_{2}$RuO$_{4}$~\cite{sigristHe3}. Such spin-triplet superconductivity
in Sr$_{2}$RuO$_{4}$ is supported by $^{17}$O NMR Knight shift
data showing that the spin susceptibility is not affected by the
superconducting state for magnetic fields parallel to the RuO$_2$
planes~\cite{Knightshift3}. Muon-spin rotation
($\mu$Sr) experiments confirm the
appearance of an internal magnetic field below the transition
temperature; thus, time-reversal symmetry is broken pointing to an
odd-parity p-wave state~\cite{TRSbreak3}.
However, the actual symmetry of the superconducting state in
Sr$_{2}$RuO$_{4}$ is still a matter of debate. Possible
f-wave superconductivity in Sr$_{2}$RuO$_{4}$ has been
suggested~\cite{f-maki3} based on evidence of a line-node
gap~\cite{node3} which does not fit with the previously proposed
isotropic (full energy gap) p-wave model, prompting further
investigation~\cite{aoki3,takimoto3}.

The above discussion suggests that it is
important to understand the mechanisms that lead to the large
quasiparticle mass enhancements~\cite{mackenzie-ARPES3} in \sruo .
One possible method
to achieve this is to compare cyclotron
resonance measurements with magnetic quantum
oscillations~\cite{1cr-mass3,2cr-mass3,3cr-mass3}.
Quantum oscillation measurements allow one to deduce quasiparticle
masses renormalized by {\it both} electron-electron and electron-phonon
interactions (the orbitally-averaged
{\it effective mass} $m^*$)~\cite{1cr-mass3,2cr-mass3,3cr-mass3}.
By contrast, a cyclotron resonance experiment
provides information about the {\it dynamical mass}
$m_{\lambda}$, which, to first order,
is expected to be affected only by
electron-phonon interactions~\cite{1cr-mass3,2cr-mass3,3cr-mass3}.
Therefore, a comparison of $m^*$ and $m_{\lambda}$ potentially
allows one to identify the relative importance of
electron-electron and electron-phonon interactions.
In this paper, we describe a measurement of cyclotron resonance
in \sruo . A resonant cavity that is able to rotate
in the magnetic field~\cite{schrama} allows a detailed
study of the evolution of resonance features with angle,
allowing unambiguous association of the resonances with
the various Fermi-surface sections.
\section{Experimental Details}
\label{sectechnique}
\begin{figure}[t]
\begin{center}
\includegraphics[height=6cm]{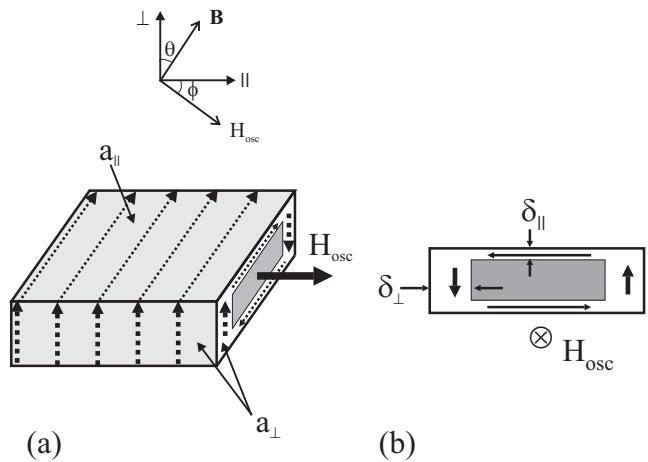}
\caption{(a) The sample is placed in the cavity with the
oscillating millimetre-wave field $H_{\rm osc}$ in the highly
conducting planes ($\parallel$ direction). (b) Both in-plane and
interplane oscillating currents $J_{\rm osc}$ are induced. In the
skin-depth regime in-plane currents flow within a surface layer of
thickness $\delta_{\parallel}$ parallel to large sample faces,
interplane currents flow within a thicker layer $\delta_{\perp}$
parallel to sample edges.} \label{expconfig}
\end{center}
\end{figure}

Cyclotron resonance in \sruo ~was observed
using a millimetre-wave cavity perturbation technique~\cite{schrama}.
Such measurements have been
shown to probe the bulk conductivity properties of
a number of anisotropic
conductors~\cite{schrama,ardavanFTR3,hill-config3,cr-bulk3}.
In the current experiment, changes in the dissipation of the
cavity ($Q$ factor) are measured as a function of an external quasistatic
magnetic field $\mathbf{B}$ with the
millimetre-wave frequency $f$ held constant~\cite{schrama}.
It is assumed that changes in the
dissipation of the cavity are due to changes in the
dissipation of the sample inside~\cite{schrama,hill-tech3}.

The sample is placed in the middle of a rectangular cavity such
that the millimetre-wave magnetic field $\mathbf{H_{\rm osc}}$ is
polarized parallel to the sample's highly conducting planes
(perpendicular to the $c$-axis) (see Figure~\ref{expconfig})~\cite{schrama}.
The response of an anisotropic conductor
such as \sruo ~in this electromagnetic environment is
understood by examining the polarizations of the currents induced
in the sample by $\mathbf{H_{\rm osc}}$. For $\mathbf{H_{\rm osc}}$
parallel to the $(a,b)$ plane (the highly-conducting plane),
both in-plane and
interlayer currents are induced (Figure~\ref{expconfig})~\cite{schrama}.
These currents flow within
a distance $\sim \delta$ (here $\delta$
is the skin
depth) from the sample's edges and faces (Figure~\ref{expconfig}~(b)).
We can estimate in-plane
($\delta_{\parallel}$) and interlayer ($\delta_{\perp}$) skin
depths in \sruo ~using the expression
$\delta=(\sigma\pi f\mu_0)^{-\frac{1}{2}}$ where $\sigma$ is the
conductivity and $f$ is the measurement frequency~\cite{bleaney3}.
At 0.7~K, $f=70.9$~GHz (the resonant frequency of the cavity~\cite{schrama}) and
$\sigma_{\rm ab}=1/\rho_{\rm ab}=1\times 10^8$~$\Omega^{-1}$m$^{-1}$~\cite{mackenzie-properties3},
where $\sigma_{\rm ab}$ is the
conductivity within the $ab$ planes;
hence, the
in-plane skin depth is
$\delta_\parallel=(\sigma_{\rm ab}(2\pi f)\mu_0/2)^{-\frac{1}{2}}=0.13~\mu$m.
The corresponding
interlayer skin depth $\delta_{\perp}$ is expected to be $\sim 3-5~\mu$m,
since
$[\sigma_{\rm ab}/\sigma_{\rm c}]^{\frac{1}{2}}\sim 30-40$~\cite{mackenzie-properties3}.

Single crystals of \sruo ~from two batches were
used for the current experiments; Sample A (lower quality)
was from batch C85A10a ($T_{\rm
c}$=1.36~K, sample size $\approx 1.0\times0.7\times~ \ltsim ~0.2$~mm$^3$),
whereas Sample B (higher quality)
was from batch C129A1 ($T_{\rm c}$=1.42~K, sample size $\approx
1.1\times0.9\times ~\ltsim ~0.1$~mm$^3$).
In the skin-depth regime, the dissipation in the
sample is governed by the surface resistance which is proportional
to the appropriate skin depth ($\delta_\parallel$ or
$\delta_{\perp}$) multiplied by an appropriate sample area
($A_{\parallel}$ or $A_{\perp}$) for the surface across which the
current travels~\cite{schrama,hill-config3,bleaney3,poole3}. The ratio of
the power dissipation due to interlayer $(P_{\perp})$ and in-plane
$(P_{\parallel})$ currents is given by
$P_{\perp}/P_{\parallel}=A_{\perp}\delta_{\perp}/A_{\parallel}\delta_\parallel$.
For Sample~B the area ratio is
$A_{\parallel}/A_{\perp}\sim 10$ and the power dissipation ratio
is $P_{\perp}/P_{\parallel}\sim 4$, which indicates that both
$\sigma_{\rm c}$ and $\sigma_{\rm ab}$ contribute to the
dissipation within the cavity. Thus
we expect to be sensitive to high-frequency effects
involving {\it both} in-plane and interlayer currents~\cite{schrama}.

The cavity, constructed out of silver rectangular waveguide, has
dimensions $6\times 3 \times 1.5$~mm and resonates in the
TE$_{102}$ mode at 70.9~GHz~\cite{schrama}.
The sample is placed in the centre of
the cavity; thus for the TE$_{102}$ mode the sample sits in the
oscillating magnetic field antinode. On introducing a typical
crystal, the quality factor of the cavity changes from $\sim$1500 to
$\sim$1200 and the resonant frequency shifts by $\sim$130~MHz. The
oscillatory electromagnetic field in the cavity induces currents
in the sample with a magnitude proportional to the sample's
conductivity~\cite{schrama,ardavanFTR3,cr-bulk3,hill-tech3}. The currents
dissipate energy, affecting the quality factor of the cavity and
thus the amplitude of the millimetre-wave transmitted through it.
Cyclotron resonance appears as a maximum in the high-frequency
conductivity; we detect it as a minimum in the transmission of the
cavity. The cavity is mounted inside a special $^3$He cryostat
(temperatures $0.48~{\rm K}\leq T \leq 10$~K)
which allows it to be rotated in the magnetic field~\cite{schrama};
the radiation travels to and from the cavity via a pair of
rectangular stainless steel waveguide. The waveguide and the
cavity are coupled using recessed circular coupling holes and two
rexolite cylinders which align the iris in the waveguide with the
cavity iris and provide a convenient rotation axis for the cavity~\cite{schrama}.
A Millimetre-wave Vector Network Analyser (MVNA) was used as a
source and  detector of the millimetre-wave radiation~\cite{schrama}.

The measurement was performed in quasistatic magnetic fields of up
to 45~T supplied by resistive and hybrid magnets at NHMFL, Florida State
University; further experiments employed a superconducting magnet
in Oxford, which provided fields of up to
17~T.
\section{Measurements and Results}
\label{secresults}
The cavity transmission was measured with the
cavity and sample at many different orientations in
the static magnetic field.
Two angles, $\phi$ and $\theta$, are used describe the
geometry of the measurements.
\begin{itemize}
\item $\phi$ describes the orientation of the sample inside the
cavity. It is the angle between the sample $a$-axis and the
direction of the oscillating magnetic field inside the cavity,
$H_{\rm osc}$. The angles $\phi= 0^\circ, 45^\circ$ were studied
for sample B ($T_{\rm c}$=1.42~K) and $\phi=
0^\circ,45^\circ,60^\circ$ for sample A ($T_{\rm c}$=1.36~K).
\item $\theta$ is defined as the angle between the external
quasi-static magnetic field $B$ and the normal to the sample
$(a,b)$ plane, {\it i.e}.\ the angle between the magnetic field
$B$ and the Fermi cylinder axes. The angle $\theta$ provides
detailed information about the evolution of cyclotron resonance
features.
\end{itemize}
Following the angle-dependent-magnetoresistance-oscillation
(AMRO)~\cite{ohmichi-AMRO3} and de Haas-van Alphen
oscillation~\cite{yoshida3,96mackenzie3,ohmichi3} experiments which have been
used to derive the Fermi-surface topology,
we expect a twofold symmetry of the cyclotron
resonance with respect to $\theta$ (negative
and positive angles) and a fourfold symmetry with respect to
$\phi$.
\begin{figure}
\includegraphics[height=8cm]{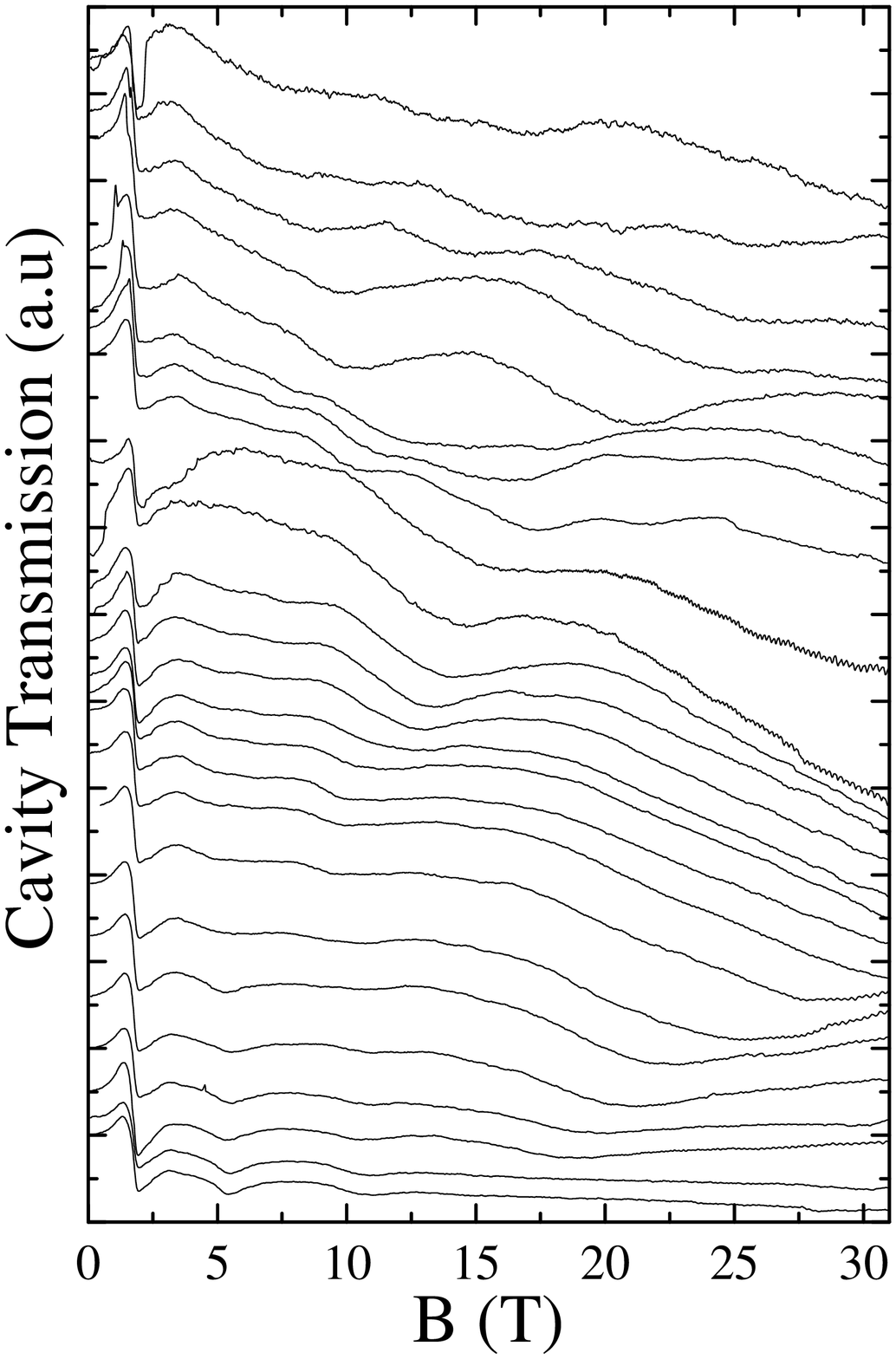}
\includegraphics[height=8cm]{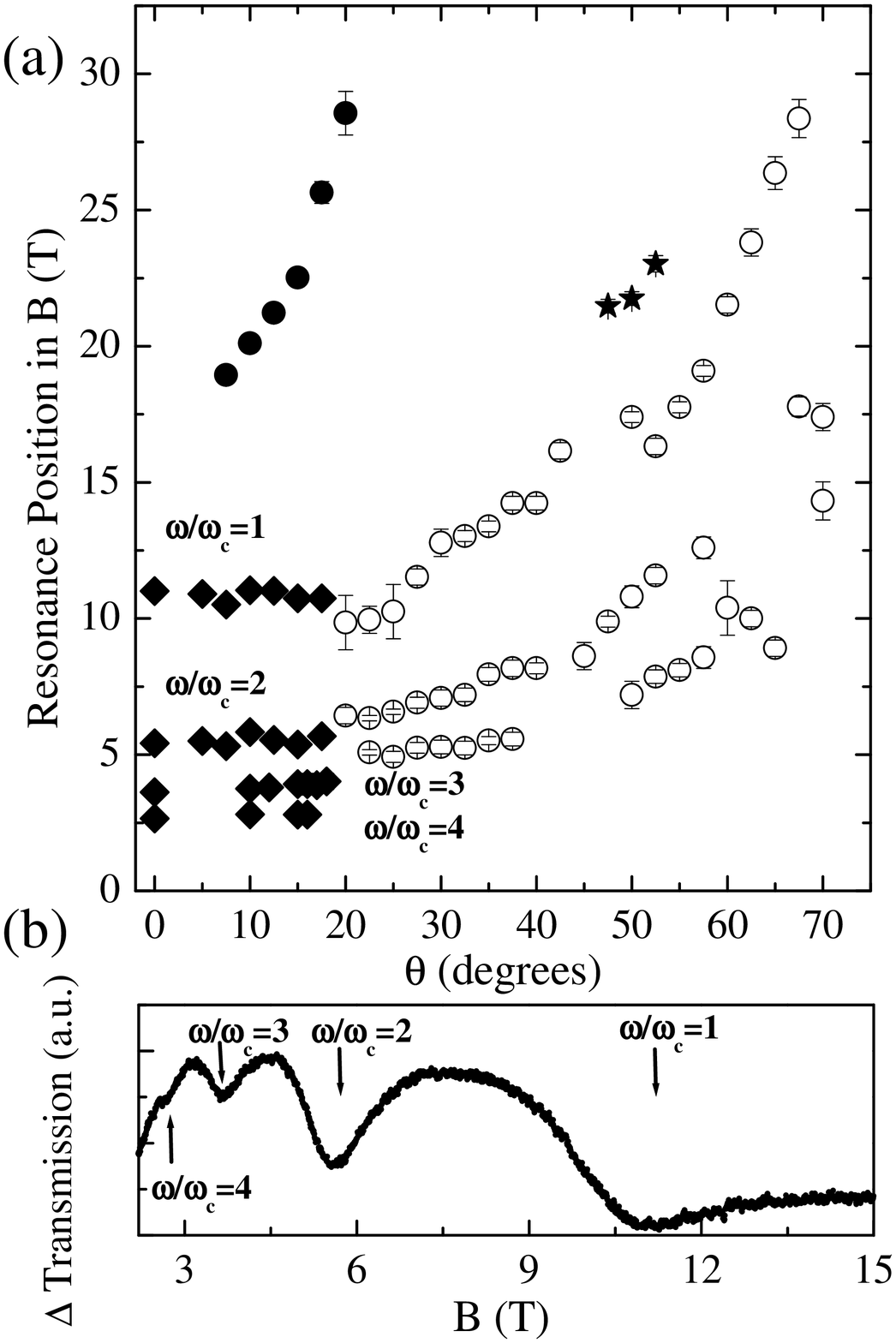}
\caption{Left: cavity transmission as a function of magnetic field for
\sruo ~Sample~B;
$\phi=0^\circ$ and $0^\circ$(bottom)$\leq\theta\leq70^\circ$(top).
The measurement frequency is $f=71.0$~GHz. The temperature is 0.6~K.
The traces are offset for clarity.
Right: (a)The positions of resonances from the left-hand figure
plotted against $\theta$. Series of resonances are labelled with
filled diamonds, circles, stars and empty circles. (b) Harmonic
series visible in the raw data at low fields and small angles
$\theta$ ($\theta=0^\circ$, corresponds to filled diamonds).}\label{raw0}
\end{figure}

The left-hand side of Figure~\ref{raw0}
shows the data collected for $\phi=0^\circ$ and
Sample~B ($T_{\rm c}$=1.42~K). Here, the
transmission of the cavity is plotted as a function of the
magnetic field for a range of field orientations, $\theta$.
Measurements were taken every 2.5$^\circ$ for
$0^\circ \leq \theta \leq 70^\circ$. Above $\theta=70^\circ$ the
signal-to-noise ratio becomes poor because
the coupling between the waves in the waveguide and the wave in
the cavity is reduced with angle by a factor ($\cos\theta$);
hence the amplitude of transmission of the whole system (signal in and
out) falls by a factor ($\cos^2\theta$) (reduction factor
$\cos^2(70^\circ)=0.12$)~\cite{schrama,ardavanFTR3}. The data are shown
offset for clarity, with no background subtraction. The
absorption visible in every trace at 2~T is due to magnetic impurities in the
stainless steel waveguide used to construct the
insert~\cite{schrama,ardavanFTR3}. This absorption is independent of
$\theta$ and of temperature. The sweeps also show quantum
oscillations; analysis of their frequency is in good agreement
with the $\alpha$ and $\beta$ frequencies reported
previously~\cite{yoshida3,96mackenzie3,ohmichi3}. Thus, we can be
confident that the millimetre waves
are well coupled to the sample within the
cavity. The detailed angle dependence of the quantum oscillations
is described in Section~\ref{secQO}.

Initially, for $\theta\leq20^\circ$, three absorption features are
clearly visible in Figure~\ref{raw0} and their evolution can be
easily tracked at small angles. For $\theta>20^\circ$ the spectrum
becomes increasingly complicated, with many additional
absorption features becoming visible. The positions in field of
the observed minima in transmission (corresponding to maxima in
the conductivity of the sample) are plotted in
the right-hand side of Figure~\ref{raw0} as a function of angle.
This visually simplifies the complex evolution of
features; for example, inset (b) of
Figure~\ref{raw0} shows one set of raw data taken for
$\theta=0^\circ$ at low field. Four absorptions can be
observed; their positions are plotted above in (a),
corresponding to the four black diamonds at $\theta=0^\circ$.

\begin{figure}
\includegraphics[height=8cm]{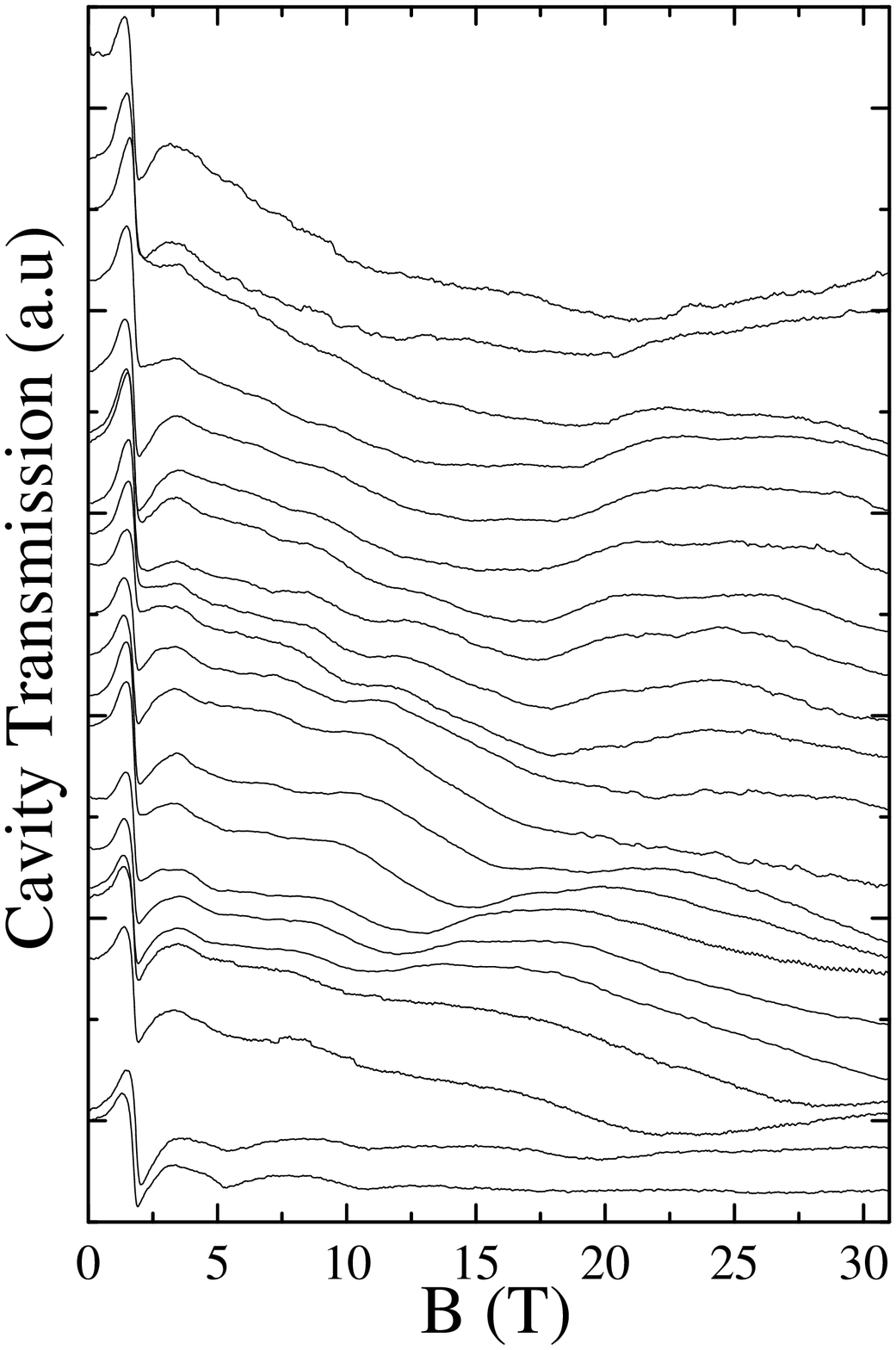}
\includegraphics[height=8cm]{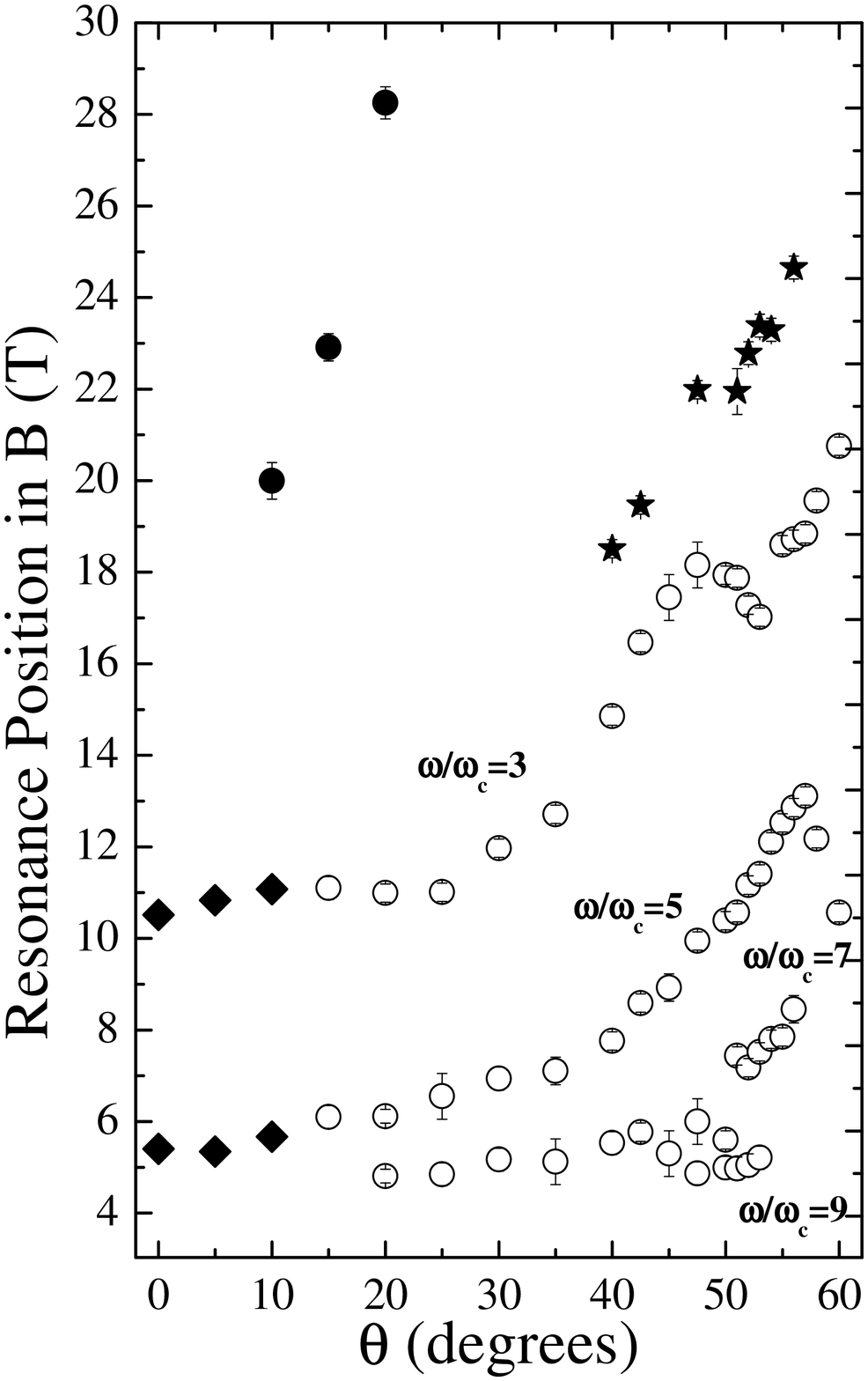}
\caption{Left: Cavity transmission as a function of magnetic field for
Sample~B;
$\phi=45^\circ$ and $0^\circ$(bottom)$\leq\theta\leq60^\circ$(top)
The measurement frequency is 70.9~GHz. The temperature is 0.6~K.
The traces are offset for clarity.
Right: the positions of the resonances in the left-hand figure
plotted against the angle $\theta$. Separate series of resonances are
grouped and labelled with filled diamonds, circles, stars and empty
circles.} \label{raw45}
\end{figure}

We have also performed a more detailed exploration of the
angle dependence of the transmission for a different orientation
of the sample within the cavity. The left-hand side
of Figure~\ref{raw45} shows the
transmission of the cavity containing Sample~B as a
function of magnetic field and angle $\theta$. The sample was
again placed in the cavity centre but its orientation was changed
so that $\phi=45^\circ$. For $\theta\ge40^\circ$ the angle
was changed in $1^\circ$ increments up to $\theta=60^\circ$.
These data and data obtained at various angles $\phi$ for the
lower quality sample, show that the spectra are $\phi$
independent, confirming the largely cylindrically symmetric nature
of the Fermi-surface sections~\cite{bergemann3}.

The absorption features evolve identically to the ones obtained
for $\phi=0^\circ$. The position of the absorption features in
field $B$ is plotted against $\theta$ in the right-hand side
of Figure~\ref{raw45}. Data
obtained from the lower quality sample ($T_{\rm c}$=1.36~K) are
also qualitatively similar, although the features are less clearly
resolved and their evolution is more difficult to trace. The good
agreement between the various measurements means that we can be
confident that the absorption features we observe and the
trends we identify in their evolution are robust and reproducible.
\section{Interpretation of data}
We shall now explain the origin of the features
identified by open circles, and black circles, squares, diamonds
and stars shown in Figure~\ref{invB0}.
This figure plots the inverse of the resonance field positions for Sample~B
as a function of $\theta$;
as will become apparent below,
this method of displaying the
data makes identification of the various series of cyclotron
resonances much easier.

Before allocating the different resonances to the
various Fermi-surface sections of \sruo ,
it is necessary to discuss the various mechanisms
that give rise to harmonics of cyclotron resonance.
\subsection{Cyclotron harmonics}
The presence of harmonics of the fundamental cyclotron
resonance frequency is most easily discussed in terms
of the semiclassical motion of quasiparticles
on the Fermi surface in a magnetic field.
This is determined by the
Lorentz force~\cite{ashcroft3},
\begin{equation}
\hbar \frac{{\rm d}{\bf k}}{{\rm d}t}=q {\bf v} \times {\bf B},
\label{lorentz}
\end{equation}
where $q$ is the quasiparticle charge and
where the
quasiparticle velocity {\bf v} is given by
\begin{equation}
\hbar {\bf v}= \nabla_{\bf k}E({\bf k}).
\label{veleqn}
\end{equation}
Here $\nabla_{\bf k}$ is the gradient operation in $k$-space
and $E({\bf k})$ is the quasiparticle energy~\cite{ashcroft3}.
Equations~\ref{lorentz} and \ref{veleqn}
produce orbits on the Fermi surface in planes
perpendicular to the magnetic field.

All three Fermi-surface sections in \sruo
~are known to be weakly corrugated cylinders~\cite{bergemann3}.
As long as {\bf B} is {\it not} applied
in the intralayer plane, all
of the field-induced orbits in \sruo ~will be closed orbits,
enabling {\it cyclotron frequencies}
$\omega_{\rm c}$ to be defined;
\begin{equation}
\omega_{\rm c}=\frac{qB}{m^*_{\rm CR}},
\end{equation}
where $m^*_{\rm CR}$ is defined by
\begin{equation}
m^*_{\rm CR}= \frac{\hbar^2}{2 \pi}\left(\frac{{\rm d}A}{{\rm d}E}\right).
\label{massdef}
\end{equation}
Here $A$ is the cross-sectional $k$-space area of the closed orbit~\cite{ashcroft3}.
In \sruo ,~the weak distortions in the interlayer ($k_z$)
direction have little effect on the cross-sectional
areas of the almost-cylindrical
Fermi-surface sections~\cite{bergemann3};
hence, the corresponding cyclotron frequencies
will not depend much on $k_{\rm z}$.

The contributions that this orbital motion makes
to the various components of the frequency-dependent
conductivity tensor are determined by the evolution
of the quasiparticle velocity~\cite{schrama,sjb-inplane3,hillPOR3}.
Oscillations in a particular velocity component
at frequencies that are harmonics of $\omega_{\rm c}$ will lead
to resonances in the corresponding
conductivity tensor component at the
same harmonics~\cite{schrama,sjb-inplane3,hillPOR3}.
We now distinguish two sources of these oscillations.
\subsubsection{Odd harmonics of $\omega_{\rm c}$ in the intralayer conductivity}
\label{intralayer}
If a cylindrical Fermi-surface section has a non-elliptical
cross-section, then the intraplane quasiparticle velocity components
$v_x$ and $v_y$ will oscillate non-sinusoidally,
the oscillation being describable by a series
of Fourier components at harmonics of $\omega_{\rm c}$~\cite{sjb-inplane3}.
When, as is the case in \sruo ,~the Fermi-surface
section has inversion symmetry~\cite{bergemann3}, only odd harmonics
are observed~\cite{sjb-inplane3}.
Therefore, resonances in the frequency-dependent
intralayer conductivity $\sigma_{zz}$
are expected at odd harmonics of $\omega_{\rm c}$.
\subsubsection{Odd and even harmonics of $\omega_{\rm c}$ in the interlayer
conductivity}
\label{interlayer}
The weak distortions of a Fermi-surface
section in the interlayer direction cause
the quasiparticle velocity to ``rock
up and down'' in the interlayer $k_z$
direction as the quasiparticle traverses a cyclotron orbit~\cite{hillPOR3}.
This will induce an oscillatory component into
the interlayer velocity $v_z$;
experiments~\cite{schrama} and calculations~\cite{hillPOR3}
have shown that both odd and even
harmonics of $\omega_{\rm c}$ are possible.
Therefore resonances in the frequency-dependent
interlayer conductivity components $\sigma_{xx}$, $\sigma_{yy}$
are expected at
integer harmonics of $\omega_{\rm c}$.
\subsection{Identification of cyclotron resonance series}
The almost-cylindrical nature of the
Fermi-surface sections of \sruo ~\cite{bergemann3} means
that the cyclotron frequency will have
the straightforward angular dependence
$\omega_{\rm c}=eB\cos\theta/m^*_{\rm CR}(0)$,
where $\theta$ it the angle between {\bf B} and
the $c$ ($k_z$) direction and $m^*_{\rm CR}(0)$ is
the cyclotron mass observed at $\theta =0$~\cite{review}.
Harmonics will therefore occur at frequencies
$\omega=neB\cos\theta/m^*_{\rm CR}(0)$, where $n$
is an integer.

\begin{figure}
\includegraphics[height=8cm]{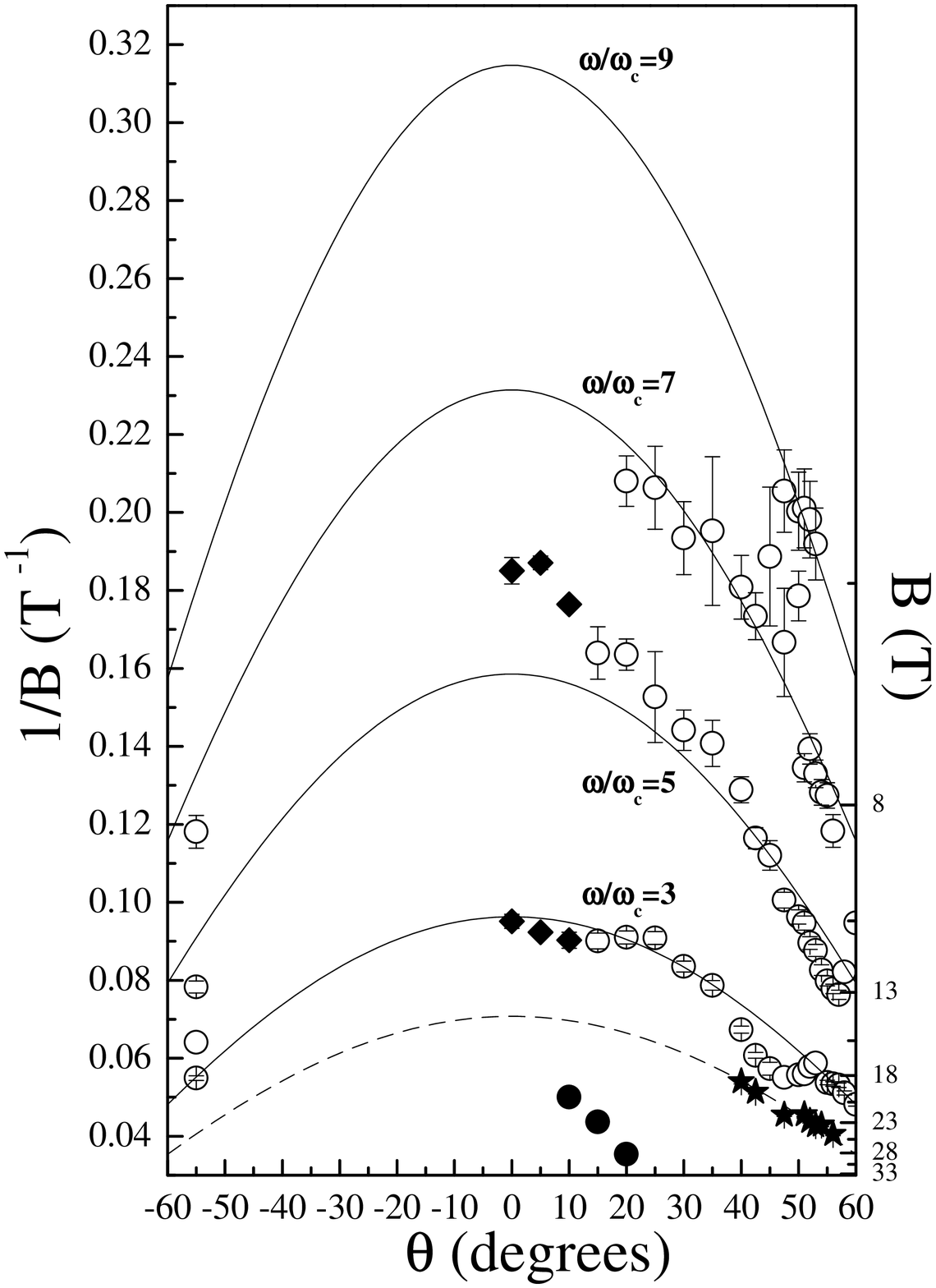}
\includegraphics[height=8cm]{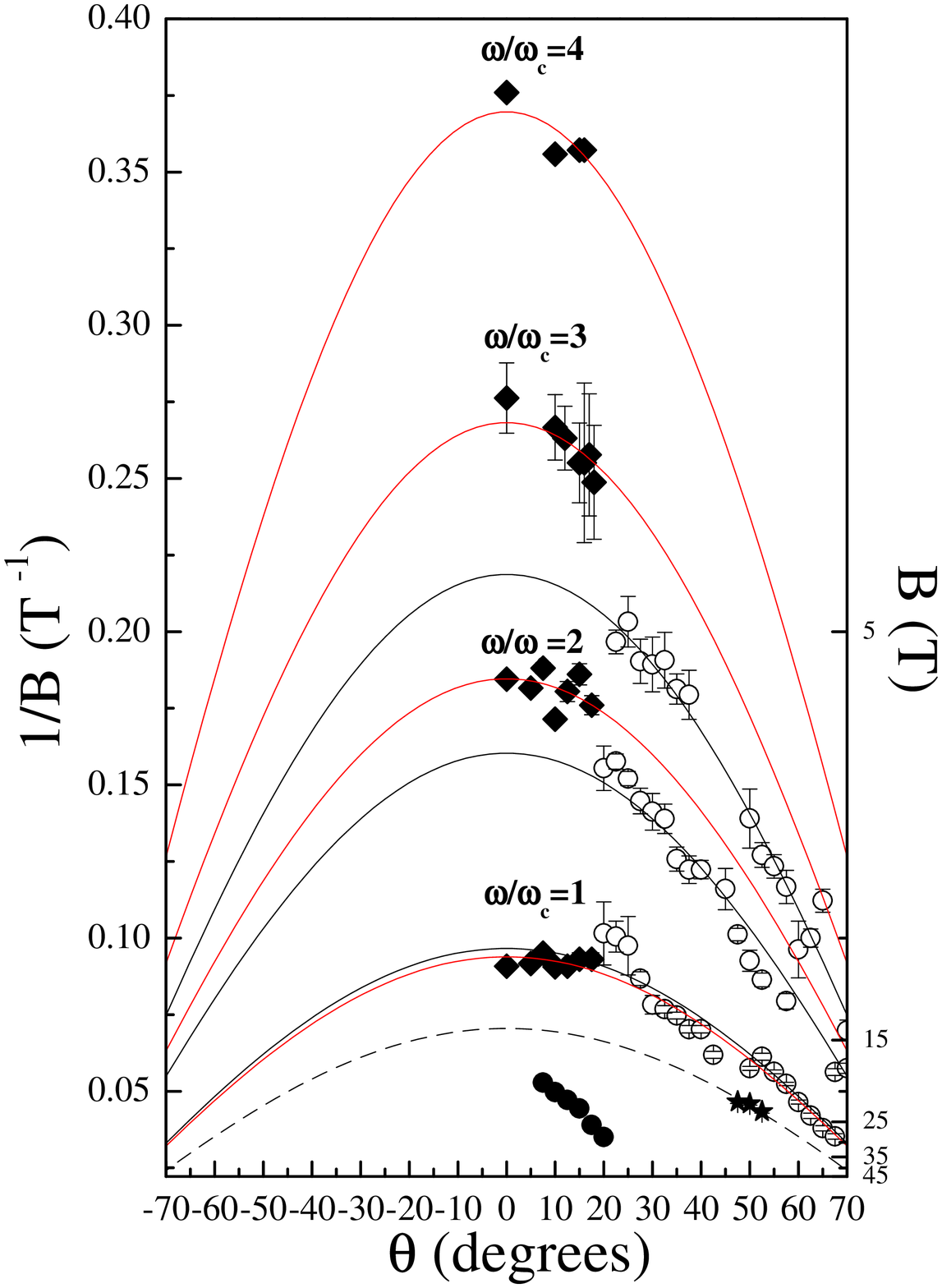}
\vspace{-0.5cm} \caption{Left: the positions of resonances from
Figure~\ref{raw45} plotted as a function of $\theta$ and the
inverse magnetic field $1/B$. Right: the positions of resonances in
Figure~\ref{raw0} plotted as a function of $\theta$ and the
inverse magnetic field $1/B$. On both sides,
the solid curves are fits of
Equation~\ref{CRfield} to the data shown as
hollow circles and filled diamonds.
Corresponding fitting
parameters $A_n$ are listed in Tables~\ref{oddharms} and ~\ref{evenharms}.
The dashed curves are fits to the data shown as stars.}
\label{invB0}
\end{figure}

In our experiment, the measurement frequency $\omega$ is fixed and
the magnetic field $B$ is swept. Absorption is
observed at particular values of $B$ defined by
\begin{equation}
B=\frac{\omega m^*_{\rm CR}(0)}{n e \cos\theta}=\frac{1}{A_n\cos \theta}.
\label{CRfield} 
\end{equation}
The harmonic branches and their cosinusoidal dependence on $\theta$
are most
easily identified when $1/B$ is plotted against $\theta$
(see Figure~\ref{invB0}); the harmonics are evenly spaced in $1/B$,
and vary as $\cos \theta$.

\begin{table}
\caption{Parameters obtained from the fits of Equation~\ref{CRfield}
 to the data shown as hollow circles in Figure~\ref{invB0}.}
\label{oddharms}
\begin{tabular}{|c|c|c|c|c|}
\hline  Fit ($T^{-1}$) & $A_3$  & $A_5$  & $A_7$ & $A_9$ \\

\hline $\phi=0^\circ$ & 0.0966 $\pm$ 0.002 & 0.160 $\pm$ 0.002 &
0.219 $\pm$ 0.002 & - \\

\hline $m_{\rm CR}(0)/m_{\rm e}$ & 4.10 $\pm$ 0.09 & 2.47 $\pm$ 0.05
& 1.81 $\pm$ 0.01 & - \\

\hline $\phi=45^\circ$ & 0.0963 $\pm$ 0.001 & 0.159 $\pm$ 0.003 &
0.231 $\pm$ 0.004 & 0.315 $\pm$ 0.003 \\

\hline $m_{\rm CR}(0)/m_{\rm e}$ & 4.11 $\pm$ 0.03 & 2.50 $\pm$ 0.05
& 1.71 $\pm$ 0.03 & 1.26 $\pm$ 0.01 \\

\hline \hline Interpretation & 3rd Harmonic & 5th Harmonic & 7th
Harmonic & 9th Harmonic \\

\hline Fermi surface & $\gamma$ & $\gamma$ & $\gamma$ & $\gamma$ \\
\hline

\end{tabular}
\end{table}
Four branches corresponding to the hollow circles are fitted by
Equation~\ref{CRfield} in
Figure~\ref{invB0};
the values of the fitting parameters $A_n$
and the corresponding cyclotron
effective masses are given in Table~\ref{oddharms}. Inspection of
the fitting parameters reveals that the ratio $A_3:A_5:A_7:A_9$ is
very close to $3:5:7:9$. This indicates that all of these
resonances arise from a single Fermi-surface section, and that we
are observing the third, fifth, seventh and ninth harmonics of the
cyclotron resonance. The observation of only odd harmonics
strongly suggests that the resonances result from
the intralayer conductivity mechanism described in Section~\ref{intralayer}.

The fundamental frequency (first harmonic) of
this series occurs at 31.2~T
for $\theta=0^\circ$, corresponding to a cyclotron mass of
$m^*_{\rm CR}(0)=12.35 m_e$ where $m_e$ is the free electron mass. Given the
high cyclotron mass, it is probable that this series originates
from the $\gamma$ Fermi-surface section, for which a thermodynamic
effective mass of $m^*_{\rm cr\gamma}=14.6 m_e$ has been deduced
from analysis of magnetic quantum
oscillations~\cite{mackenzie-ARPES3}.

Further support for this assignment can be qualitatively derived
from the attributes of the $\gamma$ Fermi-surface section~\cite{bergemann3}.
Bandstructure calculations~\cite{mazin} show
that the Fermi velocity varies strongly around the $\gamma$ sheet,
which will tend to lead to a large
contribution to the
frequency-dependent intralayer magnetoconductivity~\cite{sjb-inplane3}.
By contrast, the $\gamma$ sheet is only
very weakly warped in the interlayer direction,
leading one to expect that it only plays a small
role in the interlayer conductivity~\cite{schrama,sjb-inplane3}.\footnote{The
$\gamma$ Fermi surface accounts for only 6\% of the $c$-axis
conductivity (as opposed to the $\beta$ sheet with an estimated
86\% contribution)~\cite{bergemann3}.}

We now turn to data for
angles $\theta<20^\circ$, where the spectra are dominated by a
different set of branches (filled diamonds).
The fitting parameters for this series
of resonances are listed in Table~\ref{evenharms};
the ratios
$A_1:A_2:A_3:A_4$ are close to $1:2:3:4$ (solid line fits to
Equation~\ref{CRfield} in the right-hand side of Figure~\ref{invB0}). In
Figure~\ref{raw0}(b)we show the $\theta=0^\circ$ raw data
sweep and indicate the positions of the absorption features
corresponding to the $1^{\rm st},2^{\rm nd},3^{\rm rd},4^{\rm th}$
harmonics. The fundamental frequency (first harmonic) of this
series occurs at 10.9~T for $\theta=0^\circ$, corresponding to a
cyclotron mass of $m^*_{\rm CR}(0)=4.3\,m_e$. The fact that both even
and odd harmonics are observed is suggestive of the
interlayer conductivity mechanism for harmonics~\cite{hillPOR3} described
in Section~\ref{interlayer}.
\begin{table}
\caption{Parameters obtained from the fits of Equation~\ref{CRfield}
to the data shown as diamonds in
Figure~\ref{invB0}.}
\label{evenharms}
\begin{tabular}{|c|c|c|c|c|}
\hline  Fit ($T^{-1}$) &  $A_1$ & $A_2$ & $A_3$ & $A_4$ \\ \hline

$\phi=0^\circ$ & 0.0939 $\pm$ 0.001 & 0.185 $\pm$ 0.002 & 0.268
$\pm$ 0.002 & 0.370 $\pm$ 0.003 \\ \cline{1-5}

$m_{\rm CR}(0)/m_e$ & 4.22 $\pm$ 0.04  & 2.14 $\pm$ 0.03  & 1.48
$\pm$ 0.01 & 1.07 $\pm$ 0.01 \\

\hline $\phi=45^\circ$ & 0.0932 $\pm$ 0.001 & 0.184 $\pm$ 0.003 &
- & - \\

\hline $m_{\rm CR}(0)/m_e$ & 4.25 $\pm$ 0.05 & 2.14 $\pm$ 0.03 & - &
-
\\

\hline \hline

\hline Interpretation & 1st Harmonic & 2nd Harmonic & 3rd Harmonic
& 4th Harmonic \\ \hline

\hline Fermi surface & $\beta$ & $\beta$ & $\beta$ & $\beta$ \\
\hline
\end{tabular}
\end{table}

The $\beta$ sheet dominates the interlayer conductivity
and is of a form that will generate strong harmonic content in the
interlayer velocity~\cite{bergemann3}. We therefore
assign the integer harmonic series to cyclotron resonance of
the $\beta$ sheet.
Comparison of the cyclotron mass $m^*_{\rm CR}(0)=4.3\,m_e$ with masses
obtained from analysis of de Haas-van Alphen oscillations
$m^*_{\rm cr}=7.5\,m_e$~\cite{mackenzie-ARPES3} also supports the
assignment of this series to the $\beta$ sheet.

One further branch in Figure~\ref{invB0}, corresponding
to filled stars, is well fitted by Equation~\ref{CRfield}.
However, it does not appear to be simply related to the
other two
harmonic series. A detailed angle dependence study in the
$\phi=45^\circ$ orientation (every 1$^\circ$ at higher angles)
clearly shows this cyclotron resonance feature in
Figure~\ref{invB0} (filled stars). A fit to Equation~\ref{CRfield}
shown as a dashed line in Figure~\ref{invB0} gives
$A_1=0.0706\pm 0.001$ as the value
of the fitting parameter, which corresponds to
$B_{\rm CR}(0)=14.16$~T and a cyclotron mass of
$m^*_{\rm CR}(0)=5.60~m_e$. This feature is observed at higher angles and has
no simple relation to the other series of harmonics visible in
this region. No cyclotron resonance harmonics are observed,
and indeed it
would be difficult to argue convincingly for their presence since
they would lie very close to the other series of branches which
have a much larger absorption amplitude.

This cyclotron resonance may originate from the $\alpha$ Fermi
pocket. The $\alpha$ Fermi surface is of little importance in
interlayer transport~\cite{96mackenzie3}, and so this would be a
fundamental cyclotron resonance observed in the in-plane
conductivity. For the $\alpha$ Fermi surface an effective mass of
$m^*_{\rm cr\alpha}=3.4 m_e$ has been deduced from analysis of
magnetic quantum oscillations~\cite{mackenzie-ARPES3}.
\subsection{Summary of cyclotron masses obtained}
\begin{table}
\begin{center}
\caption{Summary of quasiparticle masses
observed in Sr$_2$RuO$_4$.
The masses from cyclotron resonance harmonics ($m_{\rm CR}(0)$)
in this work are in the fourth row.
The table also shows the corresponding
bare masses from band calculations ($m_{\rm bcr}$)
and magnetic-quantum-oscillation
experiments ($m^*_{\rm cr}$) ({\it the effective mass})
taken from reference~\cite{mackenzie-ARPES3}.
The third row shows the interpretation
of Hill~\textit{et al} of
cyclotron resonance experiments
at fixed orientation~\cite{hillCR3}.
} \label{summarytable}
\begin{tabular}{|c|c|c|c|c|}
\hline  &  $\alpha$ & $\beta$ & $\gamma$ & source \\

\hline Band Calc. ($m_{\rm bcr}/m_e$) & 1.1 & 2.0 & 2.9 &
\cite{mackenzie-ARPES3} \\

\hline Effective $m^*_{\rm cr}/m_e$ & 3.4  & 7.5 & 14.6 &
\cite{mackenzie-ARPES3} \\

\hline Cyclotron $m_{\rm CR}(0)/m_e$ & 4.33 $\pm$ 0.05 & 5.81 $\pm$
0.05 & 9.71 $\pm$ 0.2 & \cite{hillCR3}. \\

\hline Cyclotron $m_{\rm CR}(0)/m_e$ & 5.60 $\pm$ 0.03 & 4.29 $\pm$
0.05 & 12.35 $\pm$ 0.2 & This paper \\ \hline
\end{tabular}
\end{center}
\end{table}
The quasiparticle masses obtained from the various series of
cyclotron resonance harmonics ($m_{\rm CR}(0)$)
are summarised in the fourth row of
Table~\ref{summarytable}, which also shows the corresponding
({\it i.e.} orbitally-averaged)
bare masses from band calculations ($m_{\rm bcr}$)
and magnetic-quantum-oscillation
experiments ($m^*_{\rm cr}$) (the latter is usually
known as {\it the effective mass}).
In the simpler theories of many-body
effects~\cite{1cr-mass3,2cr-mass3,3cr-mass3},
$m_{\rm CR}(0)$ is closely related to
the dynamical mass
$m_{\lambda}$, which represents the
bare band mass renormalised only by electron-phonon
interactions. By contrast,
$m^*_{\rm cr}$ contains additional contributions from
electron-electron interactions.
Hence, the simple expectation is that
$m_{\rm bcr} \leq m_{\rm CR}(0) \leq m^*_{\rm cr}$~\cite{1cr-mass3,2cr-mass3,3cr-mass3}.
This does indeed seem to hold for the $\beta$ and $\gamma$
Fermi-surface sections. However,
in the case of the $\alpha$ pocket,
$m_{\rm CR}(0) > m^*_{\rm cr}$.

This variation in behaviour between the various
Fermi-surface sections is not unexpected.
A recent theory of Kanki and Yamada has shown
that, if the translational invariance of a Fermi Liquid is broken, the mass
measured in a cyclotron resonance experiment may exceed
the effective mass under some circumstances~\cite{kanki3}. The
relationship between the mass measured in a cyclotron resonance
experiment and that derived from magnetic quantum oscillations may
depend strongly on bandfilling~\cite{kanki3}.
Moreover, recent attempts
to account for the superconducting properties
of \sruo ~suggest that the many-body effects and interactions
vary amongst the Fermi-surface sections,
with, for example, some contributing
to the superconductivity and others not~\cite{f-maki3,aoki3,takimoto3}.

Finally,
Table~\ref{summarytable} compares the present results
with the cyclotron resonance experiment
of Hill~\textit{et al}.~\cite{hillCR3},
which employed an overmoded resonant cavity (sample in a
magnetic field antinode) in fields of up to 31~T at 1.4~K. They
obtained absorption spectra at several different frequencies
but at only one orientation of the sample ($\theta \approx 0$).
The features that we observe at
very small $\theta$ angles replicate Hill's measurement.
However, Hill {\it et al.} went on to assign these features
to fundamental cyclotron resonance modes
associated with the in-plane conductivity from the three Fermi
surface sections $\alpha$, $\beta$, and $\gamma$.
Our angle-dependent
study shows that this assignment (although intuitive) is
incorrect;
it is necessary to make measurements at a whole range of
sample orientations in order to observe the contributions
from all of the Fermi-surface sections.
\section{Other features in the Magneto-optical spectra}
\subsection{Large amplitude fast-moving resonance}
One resonance branch in the spectra is
difficult to assign, because it does not behave in a way
consistent with the expected angle dependence for
quasi-two-dimensional cyclotron resonance~\cite{review}
(see Equation~\ref{CRfield}).
The high-magnetic-field
low-$\theta$ feature (filled circles in Figures~\ref{raw0}, \ref{raw45}
and~\ref{invB0}), plays a dominant role in the
transmission spectra shown in Figure~\ref{raw0} and
Figure~\ref{raw45}. It is interesting that this large feature is
not observed for the very small angles $\theta=0^\circ, 5^\circ$.
Above $\theta=20^\circ$, the feature moves out of our accessible
field range.

We can make the following observations: the intensity of the
feature increases rapidly as the in-plane
component of the magnetic field grows; its amplitude then
dominates the spectra; its position shows a
rapid progression to high fields as the magnetic field
is rotated away from the $c$-axis. The first characteristic
suggests that this feature arises from a resonance in the
inter-plane conductivity. The large amplitude of the resonance
suggests that it originates from the $\beta$ Fermi surface
section, which dominates the interlayer
transport~\cite{bergemann3}. The feature is visible alongside the
odd and even harmonics we have ascribed to the variation of the
interplane velocity on the $\beta$ Fermi surface. The most
puzzling aspect of the feature is the unusual dependence of its
resonant field on angle. The cyclotron orbit period increases with
magnetic field angle $\theta$ more rapidly than $(1/\cos\theta)$,
which cannot be explained by traditional descriptions of
quasi-two-dimensional cyclotron resonance. Because of the limited
amount of data we are able to collect on the feature, the precise
dependence can not be reliably determined; even higher fields are
necessary to follow this effect.
\subsection{Magnetic Quantum Oscillations}
\label{secQO}
In quasi two-dimensional conductors magnetic quantum oscillations
have provided a
valuable method for probing the Fermi-surface
topology~\cite{bergemann3,review}. In
our experiment, quantum oscillations were observed both in the
amplitude and in the phase of the transmission signal from the
cavity containing a sample of Sr$_2$RuO$_4$. This assures us that
the millimetre waves
are well coupled to the sample and that changes in cavity signal are
due to changes in the sample properties.

Fast Fourier Transform (FFT) analysis
has been carried out on oscillations seen in the phase
and in the amplitude of the signal
transmitted by the loaded cavity
with very similar results. The oscillations in
the phase of the transmitted signal are more prominent then in
the amplitude; this is expected because the phase changes very
rapidly at resonance~\cite{schrama}
so that small shifts in conductivity produce
a large response. In this section,
we therefore concentrate on the
analysis of quantum oscillations seen in the phase of the
transmission signal from the cavity.

\begin{figure}[t]
\begin{center}
\includegraphics[height=7cm]{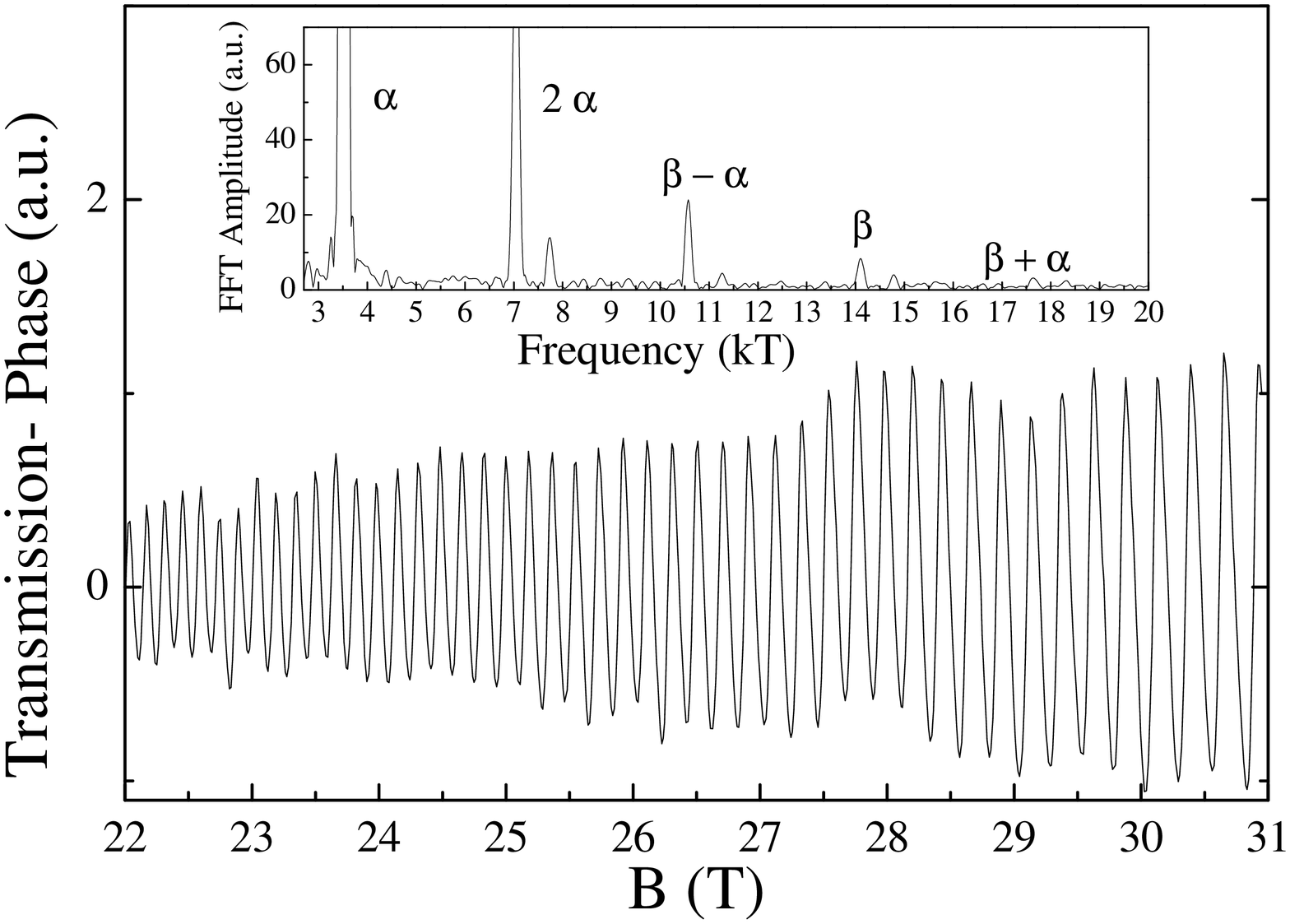}
\end{center}
\caption{Magnetic quantum oscillations
observed in the phase of the cavity transmission signal as a
function of the magnetic field $B$ at $\theta = 35^\circ$ and
temperature $T$=0.6~K. $\theta$ denotes the tilt angle
of the magnetic field with respect to the
$c$-axis. The inset shows the FFT spectrum of the
oscillations and the observed frequencies $F_\alpha$, $F_{2
\alpha}$, $F_\beta$, $F_{\beta + \alpha}$ and $F_{\beta -
\alpha}$.} \label{fftosc}
\end{figure}

\begin{figure}
\vspace{1cm}
\includegraphics[height=10cm]{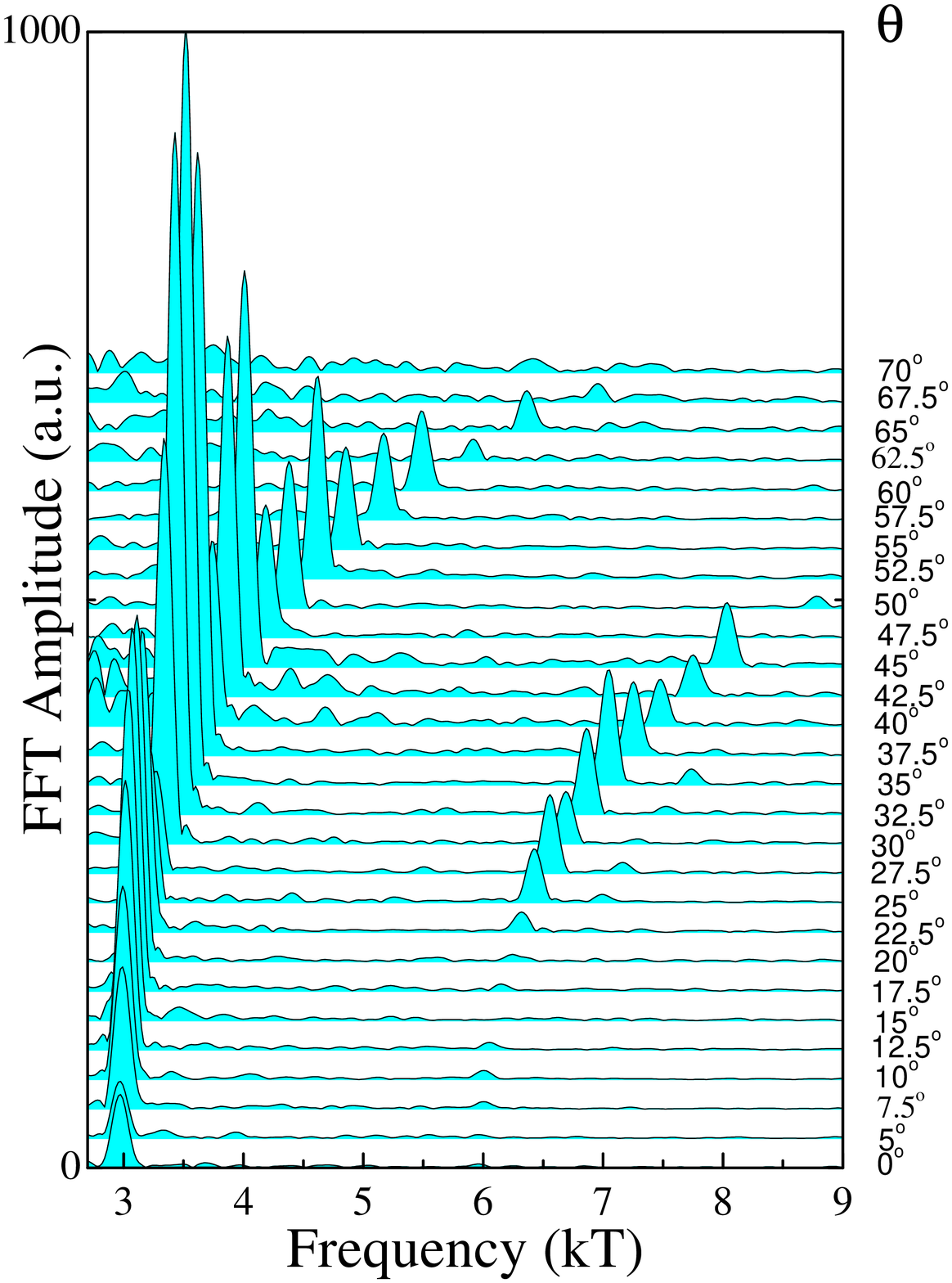}
\includegraphics[height=10cm]{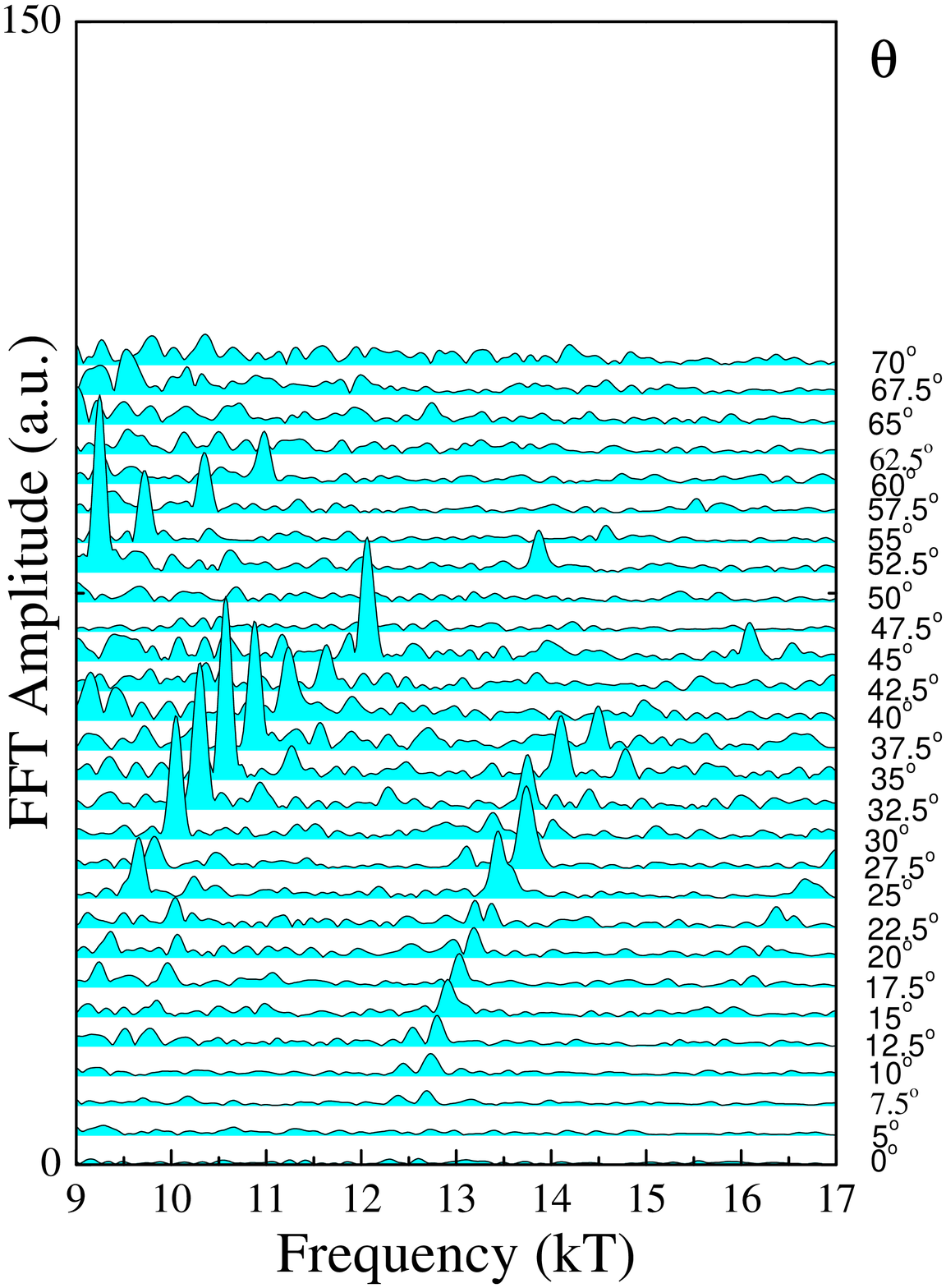}
\caption{Left: FFT spectra of quantum oscillations for measured angles
$0^\circ$(bottom)$\leq\theta\leq 70^\circ$(top) showing
frequencies around the $\alpha$ frequency and its harmonic $F_{2\alpha}$.
$T$=0.6~K.
Right:~FFT spectra of quantum oscillations for measured angles
$0^\circ$(bottom)$\leq\theta\leq 70^\circ$(top) showing
the region around the
$\beta - \alpha$, $\beta$, and $\beta + \alpha$
frequencies. $T$=0.6~K.} \label{ffthigh}
\end{figure}

Oscillatory components corresponding to the cross sectional areas
of the three Fermi-surface sections are
expected~\cite{bergemann3}. In addition, components due
to the sum or difference of the frequencies may be seen~\cite{mackenzie-ARPES3}.
The quantum oscillations in the phase of the
cavity signal show components with frequencies
$F_\alpha$ (2.88~kT), its second harmonic $F_{2 \alpha}$
(5.77~kT), $F_\beta$ (11.55~kT) and the presence of the sum and
difference frequencies $F_{\beta + \alpha}$ (8.66~kT) and
$F_{\beta - \alpha}$ (14.45~kT) shown in the $\theta=35^\circ$
sweep in Figure~\ref{fftosc}. The $F_\gamma$ frequency was not
observed because of the relatively high temperature, T = 0.7~K
(the amplitude of this frequency is very small even in sensitive
Shubnikov-de Haas and de Haas-van Alphen measurements
at 0.1~K~\cite{mackenzie-ARPES3,bergemann3}).
The evolution of the various quantum oscillation
frequencies with angle $\theta$ is shown
in Figure~\ref{ffthigh}.

\begin{figure}
\vspace{2cm}
\includegraphics[height=8cm]{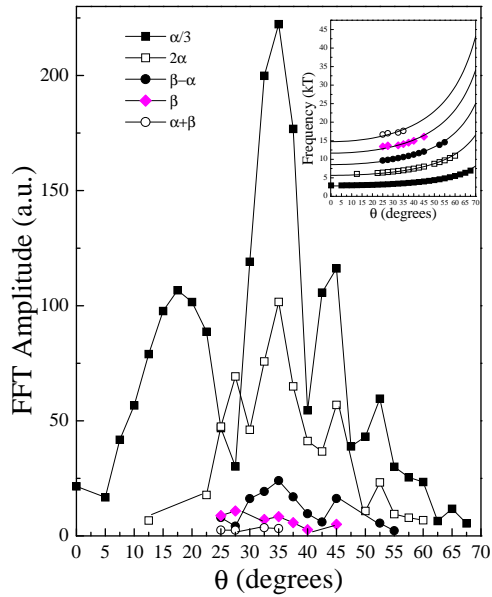}
\caption{The angle dependence of the oscillation amplitude for the
frequencies $F_\alpha$, $F_{2\alpha}$, $F_\beta$, $F_{\beta +
\alpha}$, and $F_{\beta + \alpha}$. The $F_\alpha$ amplitude is
divided by 3 in order to fit on the scale. The inset shows the
expected $F=F_{0}/\cos\theta$ evolution of observed frequencies
with angle $\theta$.} \label{fftamp}
\end{figure}

Figure~\ref{fftamp} shows the amplitude of the observed
oscillatory components as a function of the tilt angle $\theta$.
The inset of the figure shows that the frequencies of each
component shift to higher values with increasing $\theta$
according to $(1/\cos\theta)$, consistent with the angle dependence
of the cross-sectional area of the almost
cylindrical Fermi surface~\cite{bergemann3}. It is clear
that the amplitudes of all the observed frequencies are enhanced
at the angle $\theta = 35^\circ$.

In a quasi-two-dimensional
system, the amplitude of the
magnetic quantum oscillations is enhanced at the
so-called ``Yamaji angles'',
where the width of the distribution of
cross-sectional areas of
semiclassical orbits about a particular Fermi-surface section
becomes a minimum~\cite{review,yamaji3}.
In this configuration, the system behaves much
more like a purely-two-dimensional metal,
as all of the cyclotron frequencies of orbits
about the Fermi-surface section in question become very similar.
We observe this effect for
$F_\alpha$ and $F_\beta$  at $\theta = 35^\circ$, in
reasonable agreement with the maximum amplitudes of
Shubnikov-de Haas oscillations seen
at $\theta=30.6^\circ$ for the $\alpha$ frequency and
$\theta=30^\circ$ for the
$\beta$ frequency~\cite{yoshida3,ohmichi3,yoshida-dHvA3}.

The sum and difference frequencies $F_{\beta + \alpha}$ and
$F_{\beta - \alpha}$ are also enhanced at $\theta=35^\circ$.
This is because these oscillations
result mainly from the {\it chemical potential oscillation
effect} (CPOE)~\cite{ohmichi3,review,harrison3}.
In the CPOE, the chemical potential becomes pinned
to very narrow Landau levels over finite ranges of magnetic field,
modifying the way in which the Landau levels from
other Fermi-surface sections depopulate~\cite{harrison3};
hence sum and difference frequencies are generated.
The CPOE becomes most pronounced when the
Landau levels have the smallest width,
{\it i.e.} when the system looks most like a
purely two-dimensional metal~\cite{harrison3}.
Thus, the maximum effect of the CPOE
(and therefore the maximum amplitude of the resulting
mixed quantum-oscillation
frequencies) is
expected at the Yamaji angles, as observed~\cite{ohmichi3}.
\section{Summary}
\label{secsummary}
This study of the angle-dependence of the magneto-optical response
of Sr$_2$RuO$_4$ has revealed several
postulated~\cite{sjb-inplane3,hillPOR3}
cyclotron-resonance related effects, some of which have not
previously been observed.
Because of these effects,
which involve variations in the in-plane~\cite{sjb-inplane3}
and interplane~\cite{hillPOR3} quasiparticle
velocities, the magneto-optical
response of Sr$_2$RuO$_4$ is very complicated.
In order to identify the various resonances,
it is necessary to measure magneto-optical
spectra with the sample at many orientations in the magnetic
field. Our data show
that previous experiments, with the sample
at one fixed orientation~\cite{hillCR3}, resulted
in incorrect assignments and values of the cyclotron masses.

We have observed odd
cyclotron-resonance harmonics arising from the variation of the
in-plane velocities of quasiparticles on the $\gamma$ Fermi surface and
obtained a corresponding quasiparticle
mass of $m_{\rm CR}(0)=12.35 \pm 0.20 m_e$. We have also
observed odd and even harmonics arising from the variation of the
interplane velocities of quasiparticles on the $\beta$ Fermi surface;
this gives a mass of $m_{\rm CR}(0)=4.29 \pm 0.05 m_e$. Another
cyclotron resonance feature with a corresponding mass
of $m_{\rm CR}(0)=5.60\pm 0.03 m_e$
was tentatively assigned to the $\alpha$ pocket.
Finally, we
have noted an unusually strong
absorption, the field of which
evolves rapidly with sample orientation.
This feature does not appear to be simply related
to conventional cyclotron resonance.

The quasiparticle masses of the $\beta$ and $\gamma$
Fermi-surface sections measured
in our cyclotron resonance experiment are in accord
with the expectations of simple many-body theories,
in that they are considerably larger than the
bare masses predicted by bandstructure calculations,
but smaller than the effective masses deduced
from de Haas-van Alphen experiments.
In the case of the resonance attributed to the
$\alpha$ Fermi-surface pocket, the
quasiparticle mass measured
is {\it larger} than the effective mass
from the de Haas-van Alphen effect.
This may be related to recent calculations
of many-body effects, which show that the mass
measured in a cyclotron resonance experiment may exceed
the effective mass under some circumstances,
which depend on details of the bandstructure
and band filling~\cite{kanki3}.
\section{Acknowledgements}
This work is supported by EPSRC (UK).
The National High Magnetic Field Laboratory is supported by the
US Department of Energy (DoE), the National
Science Foundation and the State of Florida.
We thank Christoph Bergemann, Steve Blundell,
Neil Harrison, Stephen Hill, Albert Migliori and Andy Mackenzie
for stimulating discussions.

\section{References}


\end{document}